\documentclass[10pt,letterpaper,twocolumn]{article} 

\usepackage{ol2}
\usepackage{graphicx}
\usepackage{subfigure}
\usepackage{amsmath,amssymb}

\newcommand{\tr}{\textrm} 
\newcommand{\mb}{\mathbf} 

\begin{document}

\twocolumn[ 

\title{Electron acceleration driven by ultrashort\\ and nonparaxial radially polarized laser pulses}

\author{Vincent Marceau$^{*}$, Alexandre April, and Michel Pich{\'e}}

\address{Centre d'optique, photonique et laser, Universit{\'e} Laval\\
2375 rue de la Terrasse, Qu{\'e}bec, Qu{\'e}bec, G1V0A6, Canada\\
$^*$Corresponding author: vincent.marceau.2@ulaval.ca
}

\begin{abstract}
Exact closed-form solutions to Maxwell's equations are used to investigate the acceleration of electrons in vacuum driven by ultrashort and nonparaxial radially polarized laser pulses. We show that the threshold power above which significant acceleration takes place is greatly reduced by using a tighter focus. Moreover, electrons accelerated by tightly focused single-cycle laser pulses may reach around 80\% of the theoretical energy gain limit, about twice the value previously reported with few-cycle paraxial pulses. Our results demonstrate that the direct acceleration of electrons in vacuum is well within reach of the current laser technology.
\end{abstract}


 ] 

The advent of ultra-intense laser facilities has triggered a growing interest in laser-driven electron acceleration, opening new possibilities for the development of compact electron accelerators~\cite{malka08_naturephys}. Among the proposed laser acceleration schemes, the use of ultra-intense radially polarized laser beams in vacuum is very promising~\cite{varin06_pre,salamin06_pra,salamin07_optlett,karmakar07_lpb,fortin10_jpb,wong10_optexpress,bochkarev11_plasmaphysrep}. This scheme, also termed \emph{direct acceleration}, takes advantage of the strong longitudinal electric field at beam center to accelerate electrons to relativistic velocities along the optical axis. According to numerical simulations, clouds of electrons accelerated by radially polarized laser beams could form well collimated attosecond electron bunches~\cite{varin06_pre,karmakar07_lpb}. However, the main drawback of direct acceleration is that substantial acceleration only occurs above a high threshold power~\cite{wong11_apl}; for electrons initially at rest to reach MeV energies, laser peak powers of at least a few terawatts would be required~\cite{wong10_optexpress}.

Extensive numerical studies of electron acceleration by radially polarized laser beams have shown that reducing the pulse duration and the beam waist size generally increases the maximum energy gain~\cite{wong10_optexpress,wong11_optlett}. However, these analyses were limited by the paraxial and slowly varying envelope approximations. When the beam waist size is of the order of the laser wavelength and the pulse duration approaches the single-cycle limit, these approximations lose their validity. 

In this Letter, we provide a simple method to study the direct acceleration of electrons in the nonparaxial and ultrashort pulse regime. Under tight focusing conditions, we show that an electron initially at rest on the optical axis may be accelerated to MeV energies with laser peak powers as low as a few gigawatts. At high laser power, we demonstrate that the use of nonparaxial single-cycle pulses allows for a more efficient acceleration, reaching about 80\% of the theoretical energy gain limit, in comparison to 40\% for pulses limited by the paraxial and slowly varying envelope approximations~\cite{wong10_optexpress}. 

In order to be accurately described, ultrashort and tightly focused pulsed beams must be modeled as exact solutions to Maxwell's equations. Recently, April~\cite{april10_intech} presented a simple and complete strategy to obtain exact closed-form solutions for the electromagnetic fields of such beams. Following his approach, an isodiffracting TM$_{01}$ pulsed beam propagating in vacuum along the $z$ axis (beam waist at $z\!=\!0$) is described by the following field components in cylindrical coordinates $(r,\phi,z)$~\cite{april10_intech}:
\begin{align}
 &\!\! E_r (\mb{r},t) = \Re\bigg\{ \frac{3 E_0 \sin \tilde{2\theta}}{2\tilde{R}} \bigg( \frac{G_-^{(0)}}{\tilde{R}^2} \!-\! \frac{G_+^{(1)}}{c\tilde{R}} \!+\! \frac{G_-^{(2)}}{3c^2}\bigg) \bigg\} \ ,  \label{eq:npTM01Er}\\
 &\!\! \begin{aligned}
E_z (\mb{r},t) =  \Re\bigg\{ \frac{E_0}{\tilde{R}} \bigg[ & \frac{(3\cos^2\tilde{\theta}\!-\!1)}{\tilde{R}}  \bigg( \frac{G_-^{(0)}}{\tilde{R}} \!-\! \frac{G_+^{(1)}}{c} \bigg)  \\
&  \ \ -\! \frac{\sin^2\tilde{\theta}}{c^2} G_-^{(2)} \bigg] \bigg\}  \ ,
\end{aligned} \label{eq:npTM01Ez} \\ 
&\!\! H_\phi (\mb{r},t) = \Re\bigg\{ \frac{E_0 \sin \tilde{\theta}}{\eta_0 \tilde{R}} \bigg( \frac{G_-^{(1)}}{c\tilde{R}} \!-\! \frac{G_+^{(2)}}{c^2}\bigg) \bigg\} \ . \label{eq:npTM01Hphi}
\end{align}
Here $\Re\{\cdots\}$ denotes the real part, $c$ is the speed of light in free space, $\eta_0$ is the impedance of free space, $E_0$ is a constant amplitude, $a$ is a real positive constant called the confocal parameter (identical for all frequency components in an isodiffracting pulsed beam),  $\tilde{R}=[r^2 + (z+ja)^2]^{1/2}$, $\cos \tilde{\theta} = (z+ja)/\tilde{R} $, and $G^{(n)}_\pm = \partial^n_t [f(\tilde{t}_+)\pm f(\tilde{t}_-)]$ with $f(t) = e^{-j\phi_0}\left( 1- j \omega_0 t/s \right)^{-(s+1)}$ and $\tilde{t}_\pm = t \pm \tilde{R}/c + ja/c$. The function $f(t)$ is the inverse Fourier transform of the Poisson-like frequency spectrum of the pulse, given by~\cite{caron99_jmodoptic}
\begin{align}
F(\omega) = 2\pi e^{-j\phi_0} \left( \frac{s}{\omega_0}\right)^{s+1} \frac{\omega^s e^{-s\omega/\omega_0}}{\Gamma(s+1)}\ H(\omega) \ ,
\end{align}
where $s$ is a real positive parameter, $\phi_0$ is the constant pulse phase, $\omega_0=ck_0$ is the frequency of maximum amplitude, and $H(\omega)$ is the Heaviside step function. The fields given by Eqs.~\eqref{eq:npTM01Er}--\eqref{eq:npTM01Hphi} may be produced by focusing a collimated radially polarized input beam with a parabolic mirror of large aperture~\cite{april10_optexpress}.

The degree of paraxiality of the beam can be characterized by $k_0 a$, which is monotonically related to the beam waist size $w_0$ and Rayleigh range $z_R$ at wavelength $\lambda_0$ by $z_R = k_0 w_0^2/2 = [\sqrt{1+(k_0 a)^2} - 1]/k_0$~\cite{rodriguez-morales04_optlett}. Therefore, $k_0 a \sim 1$ for tight focusing conditions, while $k_0 a \gg 1$ for paraxial beams, in which case $z_R \approx a$.  The pulse duration $T$, which may be defined as twice the root-mean-square width of $|E_z|^2$, increases monotonically with $s$. The peak power $P_\tr{peak}$ of the pulse is found by numerically integrating the $z$ component of the Poynting vector, $S_z=E_r H_\phi$, in the transverse plane at $z=t=0$ with $\phi_0=0$. Finally, in the limit $k_0 a \gg 1$ and $s \gg 1$, Eqs.~\eqref{eq:npTM01Er}--\eqref{eq:npTM01Hphi} reduce to the fields of the paraxial TM$_{01}$ Gaussian pulse used in \cite{fortin10_jpb}. 

\begin{figure}[!t]
\centering
\includegraphics[width = \columnwidth]{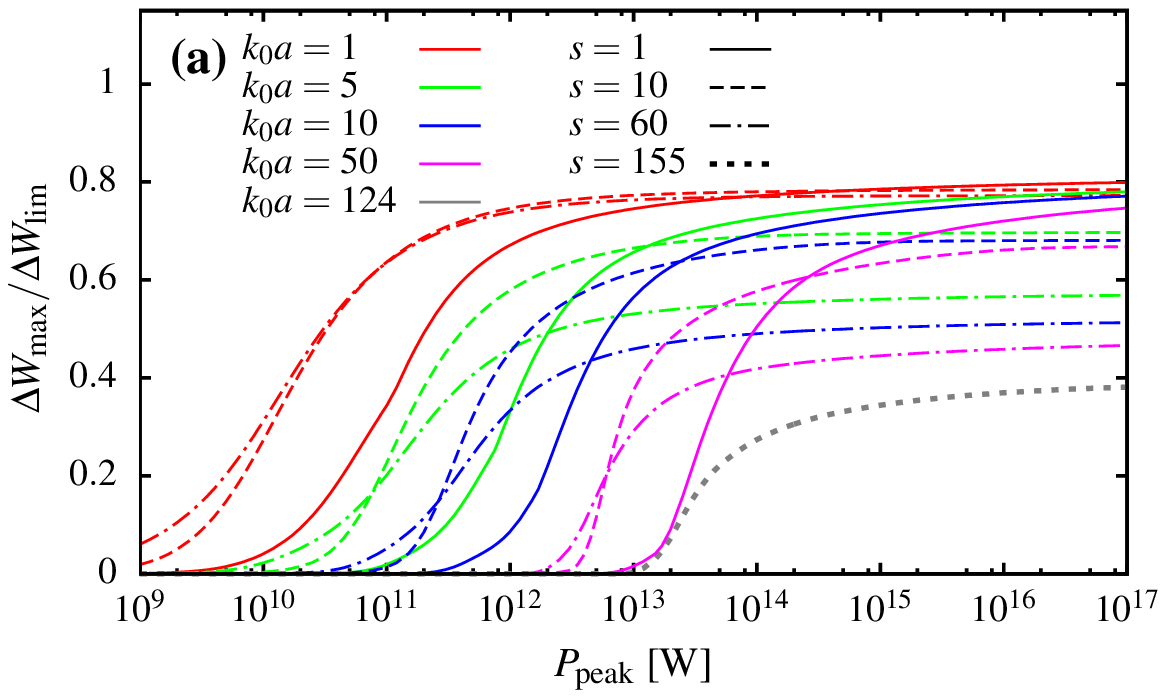} \\
\includegraphics[width = \columnwidth]{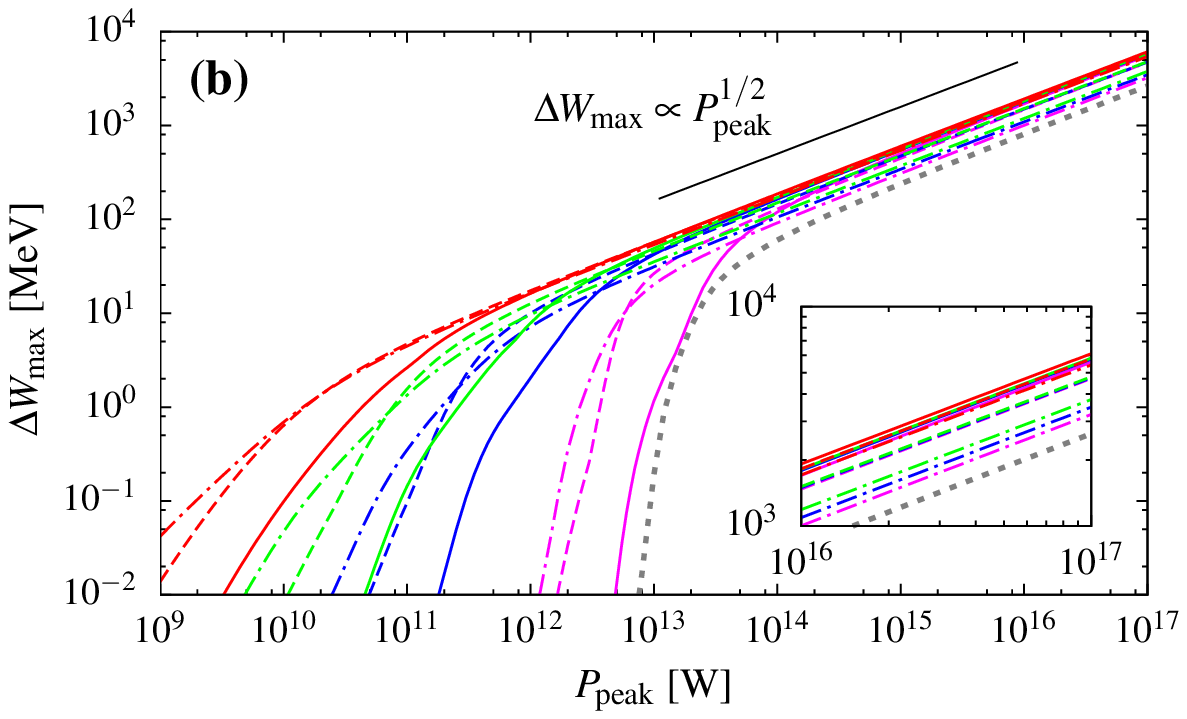} 
\caption{ (Color online) Maximum (a) normalized and (b) absolute energy gain of an electron initially at rest on the optical axis versus the laser pulse peak power for different values of $k_0 a$ and $s$. The values of $s$ used represent single-cycle ($s=1$), two-cycle ($s=10$), and five-cycle ($s=60$) pulses. The gray dashed curve ($k_0a = 124$, $s=155$) corresponds to the limit of the paraxial regime investigated in~\cite{wong10_optexpress}. The energy gain is evaluated 20 ps after the passage of the pulse at $z=0$. \label{fig:energy_power} }
\end{figure}

To simulate the laser-driven electron acceleration, the conventional Lorentz force equation~\cite{fortin10_jpb} was integrated numerically for an electron initially at rest on the optical axis outside the laser pulse at position $z_0$. Under an appropriate change of coordinates, namely $\zeta=z/a$, $\rho=r/a$, and $\tau=\omega_0 t$, it is possible to show that the dynamics is totally independent of the wavelength of maximum field amplitude $\lambda_0=2\pi/k_0$ and only depends on the parameters $k_0 a$, $s$, and $P_\tr{peak}$. For various values of $k_0 a$ and $s$, the electron initial position $z_0$ and the pulse phase $\phi_0$ were optimized to obtain the maximum energy gain $\Delta W_\tr{max}$ at different peak powers $P_\tr{peak}$. The variation of $\Delta W_\tr{max}$ with $P_\tr{peak}$ is illustrated in Fig.~\ref{fig:energy_power}. In Fig.~\ref{fig:energy_power}(a), the energy gain is normalized by the theoretical energy gain limit, $\Delta W_\tr{lim} = -e\int_0^\infty E_z(r=0,t=z/c)dz$ with $\phi_0=\pi$, which is equal to the energy gain of an electron that hypothetically remains at pulse peak from $z=0$ to infinity ~\cite{fortin10_jpb,wong10_optexpress}. For comparison, the case $w_0=2\ \mu$m and $T=7.5$ fs at $\lambda_0=800$ nm (which gives $k_0 a \approx 124$ and $s\approx155$)  is also shown in Fig.~\ref{eq:npTM01Er}. This case was previously studied in~\cite{wong10_optexpress} as the limit imposed by the paraxial and slowly varying envelope approximations.

\begin{figure}[!t]
\centering
\includegraphics[width = \columnwidth]{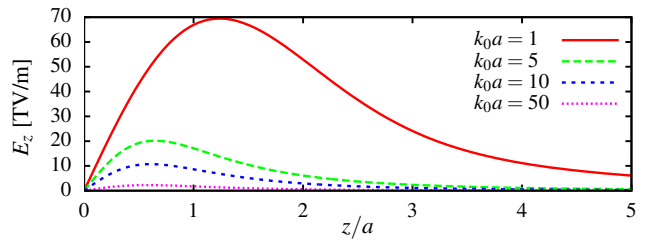}
\caption{ (Color online) Longitudinal on-axis electric field at pulse peak $t=z/c$ (computed for $\lambda_0=800$ nm) versus distance from beam waist for different values of $k_0a$. The other parameters are $\phi_0 \!=\! 0$, $s\!=\!10$, and $P_\tr{peak}\!=\!10^{12}$ W. \label{fig:Ez_s10} }
\end{figure}

At constant pulse duration, Fig.~\ref{fig:energy_power}(a) shows that the threshold laser peak power above which significant acceleration takes place is dramatically reduced as the beam focus is made tighter, i.e., as $k_0 a$ decreases. This can be attributed to the fact that the amplitude of the longitudinal on-axis electric field component increases as $k_0 a$ decreases (see Fig.~\ref{fig:Ez_s10}). Figure~\ref{fig:energy_power} also shows that longer pulses have a lower acceleration threshold and allow for higher energy gains at low peak powers. This agrees with~\cite{wong10_optexpress}, where the same phenomenon was reported to a weaker extent. This could be explained by the fact that the first cycles of the pulse preaccelerate the electron before it reaches the pulse peak, thus lowering the required threshold power. According to Fig.~\ref{fig:energy_power}(b), MeV energies may be reached at peak powers as low as 10 gigawatts with tightly focused few-cycle pulses. In comparison, a peak power of about 10 terawatts is required to reach the same energy gain with paraxial pulses. From a practical standpoint, it is however important to notice that while decreasing the acceleration threshold power, the use of tightly focused pulsed beams may reduce the number of fast electrons that are produced. The influence of $k_0a$ and $s$ on the total accelerated charge in three-dimensional simulations is the object of ongoing research.

\begin{figure}[!t]
\centering
\includegraphics[width = \columnwidth]{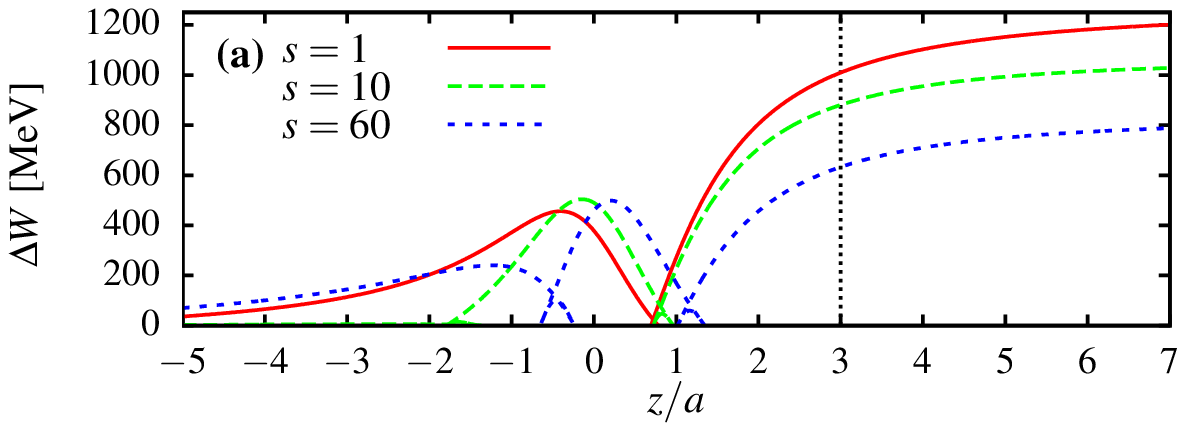}\\
\includegraphics[width = \columnwidth]{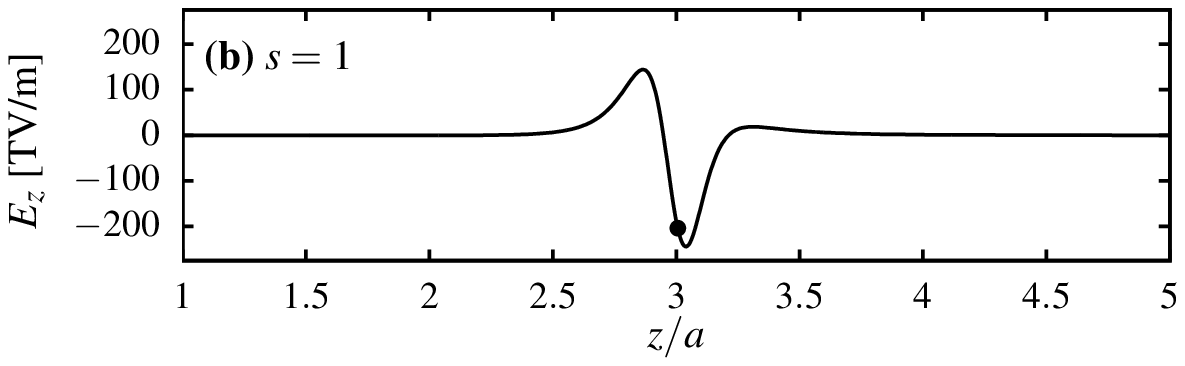}\\
\includegraphics[width = \columnwidth]{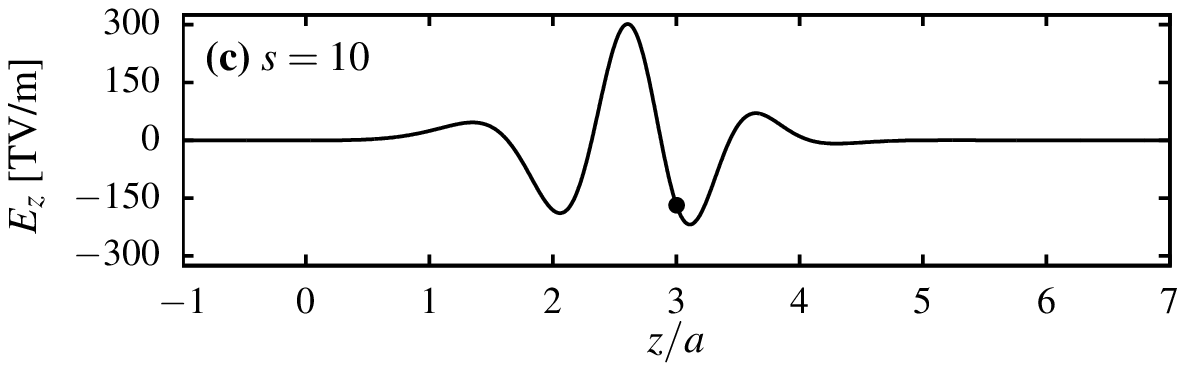}\\
\includegraphics[width = \columnwidth]{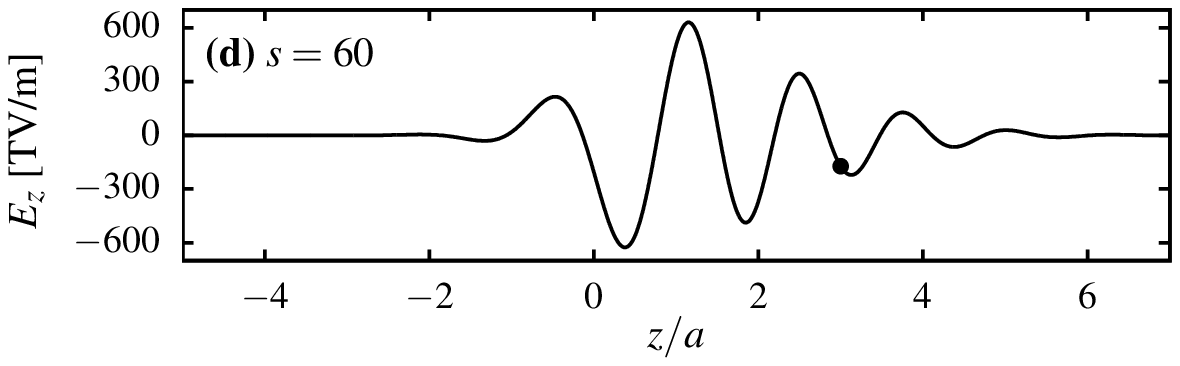}
\caption{ (Color online) (a) Energy gain versus distance from beam waist for an electron accelerated by pulses of various durations with $k_0a = 5$, $P_\tr{peak}=5$ PW. For each pulse, a snapshot of the longitudinal electric field $E_z$ (computed for $\lambda_0=800$ nm) was taken at the position indicated by the dotted vertical line and is shown in (b)--(d). The filled circle in (b)--(d) indicates the position of the electron on the optical axis.  The $z_0$ and $\phi_0$ parameters are chosen to obtain the maximum energy gain with each pulse. \label{fig:fieldprofile} }
\end{figure}

At higher peak powers, the use of shorter laser pulses yields a more efficient acceleration. Indeed, Fig.~\ref{fig:energy_power}(a) shows that electrons accelerated by a tightly focused single-cycle ($s=1$) pulse may reach 80\% of the theoretical energy gain limit, compared to 40\% with a paraxial few-cycle pulse. Shorter pulses lead to higher maximum energy gains for two reasons. First, we see in Fig.~\ref{fig:fieldprofile}(a) that with a shorter pulse, the electron enters its final accelerating cycle closer (in terms of $a$) to the focus, thus increasing the energy it is able to extract from the electric field. Second, as it is shown in Figs.~\ref{fig:fieldprofile}(b)--(d), a pulse of shorter duration allows the electron to move closer to the pulse peak during its final accelerating cycle; with longer durations, the electron is instead accelerated by the front edge of the pulse. Figure~\ref{fig:energy_power}(b) also shows that reducing the pulse duration and using a tighter focus increase the absolute value (in MeVs) of the maximum energy gain at high peak powers. In the high power regime, $\Delta W_\tr{max}$ scales as $P_\tr{peak}^{1/2}$, in agreement with results obtained under paraxial conditions~\cite{salamin06_pra,fortin10_jpb,wong10_optexpress}. 

In summary, our results highlight the importance of going beyond the paraxial and slowly varying envelope approximations in the analysis of electron acceleration in vacuum by radially polarized laser beams. We have provided a simple method to investigate the acceleration dynamics in the nonparaxial and ultrashort pulse regime by using exact closed-form solutions to Maxwell's equations. With tightly focused single-cycle laser pulses, electrons may reach around 80\% of the theoretical energy gain limit, twice the value obtained with paraxial few-cycle pulses. We have also demonstrated the possibility of significantly reducing the acceleration threshold power by using a tighter focus. According to our results, it would be possible, under tight focusing conditions such as those obtained with a high-aperture parabolic mirror, to reach MeV energy gains with laser peak powers of a few gigawatts. This is about $10^3$ times lower than the peak power required to reach the same energy gain in the limit of the paraxial approximation.  Besides being much more common in laboratories, gigawatt lasers can be operated at higher repetition rates than terawatt lasers and allow for easier pulse compression. Direct acceleration of electrons in vacuum by radially polarized laser beams is therefore much more accessible to the current laser technology than previously expected; its experimental realization is currently under investigation~\cite{payeur}.

This research was supported by the Natural Sciences and Engineering Research Council of Canada (NSERC), Le Fonds de Recherche du Qu{\'e}bec -- Nature et Technologies (FRQNT), and the Canadian Institute for Photonic Innovations (CIPI). The authors are also grateful to the CLUMEQ consortium for computational resources.



\begin{thebibliography}{10}
\newcommand{\enquote}[1]{``#1''}

\bibitem{malka08_naturephys}
V.~Malka, J.~Faure, Y.~A. Gauduel, E.~Lefebvre, A.~Rousse, and K.~Y. Phuoc,
  Nat. Phys. \textbf{4}, 447 (2008).

\bibitem{varin06_pre}
C.~Varin and M.~Pich{\'e}, Phys. Rev. E \textbf{74}, 045602(R) (2006).

\bibitem{salamin06_pra}
Y.~I. Salamin, Phys. Rev. A \textbf{73}, 043402 (2006).

\bibitem{salamin07_optlett}
Y.~I. Salamin, Opt. Lett. \textbf{32}, 90 (2007).

\bibitem{karmakar07_lpb}
A.~Karmakar and A.~Pukhov, Laser Part. Beams \textbf{25}, 371 (2007).

\bibitem{fortin10_jpb}
P.-L. Fortin, M.~Pich{\'e}, and C.~Varin, J. Phys. B: At. Mol. Opt. Phys.
  \textbf{43}, 025401 (2010).

\bibitem{wong10_optexpress}
L.~J. Wong and F.~X. K{\"a}rtner, Opt. Express \textbf{18}, 25035 (2010).

\bibitem{bochkarev11_plasmaphysrep}
S.~G. Bochkarev, K.~I. Popov, and V.~Y. Bychenkov, Plasma Phys. Rep.
  \textbf{37}, 603 (2011).

\bibitem{wong11_apl}
L.~J. Wong and F.~X. K{\"a}rtner, Appl. Phys. Lett. \textbf{99}, 211101 (2011).

\bibitem{wong11_optlett}
L.~J. Wong and F.~X. K{\"a}rtner, Opt. Lett. \textbf{36}, 957 (2011).

\bibitem{april10_intech}
A.~April, in \emph{Coherence and Ultrashort Pulse Laser Emission}, F.~J.
  Duarte, ed. (InTech, 2010), pp. 355--382.

\bibitem{caron99_jmodoptic}
C.~F.~R. Caron and R.~M. Potvliege, J. Mod. Opt. \textbf{46}, 1881 (1999).

\bibitem{april10_optexpress}
A.~April and M.~Pich{\'e}, Opt. Express \textbf{18}, 22128 (2010).

\bibitem{rodriguez-morales04_optlett}
G.~Rodr{\'i}guez-Morales and S.~Ch{\'a}vez-Cerda, Opt. Lett. \textbf{29}, 430
  (2004).

\bibitem{payeur}
S.~Payeur, INRS-EMT, 1650 Lionel-Boulet, Varennes (QC), J3X1S2, Canada (personal communication, 2012).

\end{thebibliography}
\end{document}